\documentclass[12pt,preprint]{aastex}

\slugcomment{ }

\shorttitle{The black hole fundamental plane}
\shortauthors{Wang, Wu & Kong}

\begin{document}


\title{The black hole fundamental plane from a uniform sample of radio and X-ray emitting broad line AGNs}

\author{Ran Wang and Xue-Bing Wu}
\affil{Astronomy Department, Peking University,
    Beijing 100871, China}
\email{littlestar@pku.edu.cn; wuxb@bac.pku.edu.cn}
\author{Min-Zhi Kong}
\affil{National Astronomical Observatories, Chinese Academy of
Sciences, Beijing 100012, China} \email{kmz@bao.ac.cn}

\begin{abstract}
We derived the black hole fundamental plane relationship among the
1.4GHz radio luminosity (L$_r$), 0.1-2.4keV X-ray luminosity
(L$_X$), and black hole mass (M) from a uniform broad line SDSS
AGN sample including both radio loud and radio quiet X-ray emitting
sources. We
found in our sample that the fundamental plane relation has a very
weak dependence on the black hole mass, and a tight correlation
also exists between the Eddington luminosity scaled X-ray and radio
luminosities for the radio quiet subsample. Additionally, we noticed
that the radio quiet and radio loud AGNs have different power-law slopes in
the radio--X-ray non-linear relationship. The radio loud sample
displays a slope of 1.39, which seems  consistent with the
jet dominated X-ray model. However, it may also be partly due to the
relativistic beaming effect. For radio quiet sample the
slope of the radio--X-ray relationship is about 0.85, which is possibly
consistent with the theoretical prediction from the accretion flow
dominated X-ray model. We briefly discuss the reason why our derived
relationship is different from some previous works and expect the future
spectral studies in radio and X-ray bands on individual sources in our
sample to confirm our result.
\end{abstract}


\keywords{accretion, accretion disks --- galaxies: active ---
galaxies: nuclei --- radio continuum: galaxies --- X-ray:
galaxies}

\section{Introduction}

In black hole systems, the central black hole accretion process is
believed to be the dominated energy producing mechanism,
sometimes also accompanied with a relativistic jet. Observationally,
such kinds of disk-jet systems display similar characteristics of strong
X-ray and radio emissions, and exist at different scales from the
stellar mass black hole X-ray binaries (XRBs) to  active
galactic nuclei (AGNs). Recently the non-linear relationship among
the central X-ray emission, core radio emission and black hole
mass, also called as the black hole fundamental plane, has been
investigated in details both theoretically and observationally
(eg. Merloni et al. 2003, Heinz \& Sunyaev 2003, Falcke et al.
2004). This relationship may directly reflect the common physics of
a disk-jet system around the black hole.

The radio--X-ray correlation has been widely studied in the Galactic black
hole candidates \citep{fgj03, gfp03, yc05} and has been also extended
to AGNs \citep{mhd03, fkm04}. \citet{gfp03} investigated a sample of
ten low/hard state Galaxy black hole X-ray binaries and found that all
sources in the low/hard state follow a universal power-law relation
between the radio and X-ray emissions with a  slope of
0.7 (i.e. $L_{R} \propto L_{X}^{0.7}$). Additionally, they
suggested that when the system enters the hard to soft transition
state the jet is suppressed and the radio emission drops.
\citet{mhd03} collected a large sample containing both
 Galactic black hole X-ray binaries and AGNs,
and gave a fundamental plane relation among the radio luminosity
(L$_{R\, (5GHz)}$), X-ray luminosity (L$_{X\, (2-10keV)}$) and
black hole mass (M$_{BH}$). Their result can be expressed
as $L_{R\, (5GHz)}\propto L_{X\, (2-10keV)}^{0.6} M_{BH}^{0.78}$
and the power-law slope of the radio--X-ray relation is consistent with that
previously obtained for stellar black hole
systems \citep{gfp03}.

Physics explanations on the derived radio--X ray correlation have
been also widely discussed. The radio emission is always believed to
be the synchrotron radiation from the jet while the X-ray emission
can come from both the accretion flow and the relativistic jet in a
disk-jet system. At high accretion rate the X-ray emission is mainly
from accretion flow and its dependence on black hole mass and
accretion rate can be different according to different accretion
disk model \citep{ss73, ny94, mf02}. \citet{mhd03} has investigated
the cases of several different disk models and found a good
agreement between their fitting result and the radiation inefficient
accretion flow models. When the accretion rate drops, the X-ray
emission declines according to these models. Below a certain
critical value ($L/L_{Edd}\sim10^{-5}-10^{-6}$, where $L_{Edd}$ is
the Eddington luminosity), the jet emission will become dominant
\citep{gfp03, yc05}. In such a 'low' state where jet dominates the
X-ray emission, the power-law slope of the radio--X-ray relation
 should be greater than 1 \citep{h04, yc05}. This means that when the
X-ray emission is dominated by jet, the slope of the fundamental plane
relation is probably different from the case when the X-ray emission is
dominated by the accretion flow.

However, the reliability of the fundamental plane relation given by
\citet{mhd03} has been questioned. Recently \citet{b05} pointed out
that the fundamental plane relation in \citet{mhd03} can be led out
even with those sources having only upper limit of radio emission  or
by scrambling the radio fluxes and making them randomly assigned to
the sources in the whole sample. In addition, the sample in
\citet{mhd03} has the ratios of X-ray luminosity to Eddington
luminosity in a large range, from lower than $10^{-6}$ to 1. So the
X-ray luminosity may have different dominated mechanisms according
to different level of luminosities and thus has different
dependences on the accretion rate according to the accretion disk
models. This may cause significant scatters of the intrinsic black
hole fundamental plane relation. Furthermore, the sample in
\citet{mhd03} consists of both Galactic and extragalactic sources
with estimated black hole mass and is therefore not uniform. Various
selection effects may also seriously affect the derived black hole
fundamental plan relation.

In this paper we select a uniform sample of broad line AGNs based
on the cross-identifications of ROSAT all-sky survey, Sloan Digital
Sky Survey and FIRST 20cm radio survey. With this sample we can
possibly avoid the bias from both the inhomogeneous source selections and
the data quality differences, and then better study the  black hole
fundamental plane relation for broad line AGNs. The sample
selection and data reduction process are described in section 2
and statistical results are given in section 3. The possible
theoretical explanations are discussed in section 4.

Through out this paper, we adopt the cosmology model with $H_{0} =
70\,kms^{-1}Mpc^{-1}$, $\Omega _{\Lambda} = 0.7$, and $\Omega _{M}
= 0.3$.

\section{The Sample}

\subsection{Sample selection}

Our sample is selected from the X-ray emitting SDSS AGN catalog
\citep{a03} and  the FIRST 20 cm radio survey \citep{white97,
becker03}. The AGN catalog of \citet{a03} was selected based on the
cross-identifications of the ROSAT all-sky survey (RASS) and the
Sloan Digital Sky Survey (SDSS). As discussed by \citet{a03}, the
RASS and SDSS are extremely well-matched both in the detection area
and sensitivity. With the additional information from the FIRST
radio observations, about $10^4$ RASS X-ray sources are searched for
the SDSS optical counterparts. As an initial and important result,
\citet{a03} obtained a uniform sample of more than 1200 X-ray
emitting quasars and other AGNs over the 1400 deg$^2$ of the sky,
including 964 broad permitted line AGNs (FWHM$>$1000km/s), 216
narrow permitted line AGNs (FWHM$<$1000km/s) and 45 BL Lac
candidates.

One of the benefits of this sample, as \citet{a03} pointed out, is
that the RASS and SDSS survey area is also covered by the FIRST 20cm
radio survey. So we select all the FIRST radio detected sources
in the broad line AGN catalog  and finally
construct a ROSAT-SDSS-FIRST cross-identified sample of 132 broad
permitted line AGNs. All of these 132 sources have 0.1 - 2.4keV
X-ray measurement from RASS and 1.4GHz radio measurement from the
FIRST 20cm survey. The optical spectra of these sources are available 
from the SDSS data archive, which can be used to estimate
the central black hole masses of them. For this purpose we
analyze the SDSS spectra and the data reduction process is
described in the following section.

\subsection{Data reduction and black hole mass estimation}

We examined all of the 132 spectra first with an aim that we can
finally estimate the black hole mass of AGNs with broad permitted
lines (eg. $H\beta \lambda4861, MgII\lambda2798$). We excluded the
spectra in poor quality (with no visible emission lines) or with
no H$\beta$ or MgII $\lambda$2798 lines in the observation
wavelength. At last we get the final sample of 115 objects with
black hole mass estimatable from either H$\beta\,\lambda 4861$ or
MgII $\lambda$2798 broad emission line \citep{k00,mj02,w04}. One
thing should be mentioned is that there are four high redshift
($z>2$) sources with only the CIV $\lambda$1549 broad emission
line available for black hole mass estimation in their SDSS
spectra. Because we want to reduce the scatters in
black hole mass introduced by different mass estimation methods
and the number of these high redshift sources is rather small, we exclude
these four source in our analysis as well.

We made the corrections for the Galactic extinction and redshift
effects and subtracted the iron emission from the continuum. The UV
and optical iron templates were adopted from \citet{vw01} and
\citet{bg92} in the wavelength range 1250$\AA < \lambda
_{rest}<3100\AA$ and 4250$\AA < \lambda _{rest}<7000\AA$, respectively.

Then we fit the continuum and emission lines with the {\it Mpfit}
package in IDL \footnote{http://cow.physics.wisc.edu/~craigm/idl/idl.html}, which was developed
based on the Levenberg-Marquardt technique. The continuum fitting
process was performed in the two Iron template windows respectively.
Then we fit each band of the continuum with a power-law and
calculate the rest-frame continuum flux density at $3000\AA$, and
$4400\AA$. For high redshift sources, the rest-frame optical band
moves out of the SDSS window and the $4400\AA$ flux density cannot
be fitted directly. So we extrapolate the power law fitting in the
UV band to the optical band and calculate the $4400\AA$ flux density
for these sources. Finally we fit the emission lines with the
gaussian profiles. We apply one gaussian component for MgII $\lambda
2798$ line and two gaussians for the broad and narrow components for
H$\beta\,\lambda4861$. From the fittings we can get the values of
flux and FWHM of these emission lines.

The $3000\AA$, $4400\AA$, and H$\beta$ luminosities, as well as the
0.1-2.4keV X-ray luminosity with data from RASS and rest frame
1.4GHz radio luminosity with data from FIRST, are calculated for
these 115 AGNs. We also derive their rest-frame 5GHz flux from the
1.4GHz data by assuming a power-law index of 0.5 (i.e. $f_{\nu}
\propto \nu ^{-0.5}$). Then the radio loudness is derived with the
rest frame $4400\AA$ and 5GHz flux density according to the
definition $R=f_{5GHz}/f_b$ \citep{k89}. The radio loud and radio
quiet sources are divided  by $R=10$. Our sample consists of 39
radio quiet sources and 76 radio loud sources.

Finally, the black hole masses for all of these 115 sources are
estimated. For the broad line AGNs, the black hole mass can be
estimated with the velocity and radius of the broad line region
(BLR) using the formula in \citet{k00},
\begin{equation}
M=1.464\times10^5(\frac{R_{BLR}}{light-days})(\frac{V_{FWHM}}{10^3kms^{-1}})^2M_\odot ,
\end{equation}
where $V_{FWHM}$ is the full width at half maximum (FWHM) of the
broad emission lines and $R_{BLR}$ is the radius of the broad line
region.

For the sources with H$\beta\,\lambda4861$ line measured, $V_{FWHM}$ is
the FWHM of the broad component of H$\beta$ emission line and
$R_{BLR}$ can be determined using the broad H$\beta$ line
luminosity ($L_{H\beta}$) with the empirical relation provided by
\citet{w04}(see also \cite{k05}).
\begin{equation}
{\rm Log} R_{BLR}(light-days)=(1.381\pm0.080)+(0.684\pm0.106){\rm Log}
 (L_{H\beta}/10^{42}ergs^{-1}) .
\end{equation}

For the sources with only MgII $\lambda2798$ line available, we adopt
the empirical relation provided by \citet{mj02}:
\begin{equation}
M=3.37(\frac{\lambda
L_{3000}}{10^{44}ergs^{-1}})^{0.47}(\frac{V_{FWHM,MgII}}{kms^{-1}})^2M_{\odot},
\end{equation}
where $\lambda L_{3000}$ is the $3000\AA$ continuum luminosity and
$V_{FWHM,MgII}$ is the FWHM of the MgII emission line.

\subsection{The sample properties}

\begin{figure}
\centering
\plotone{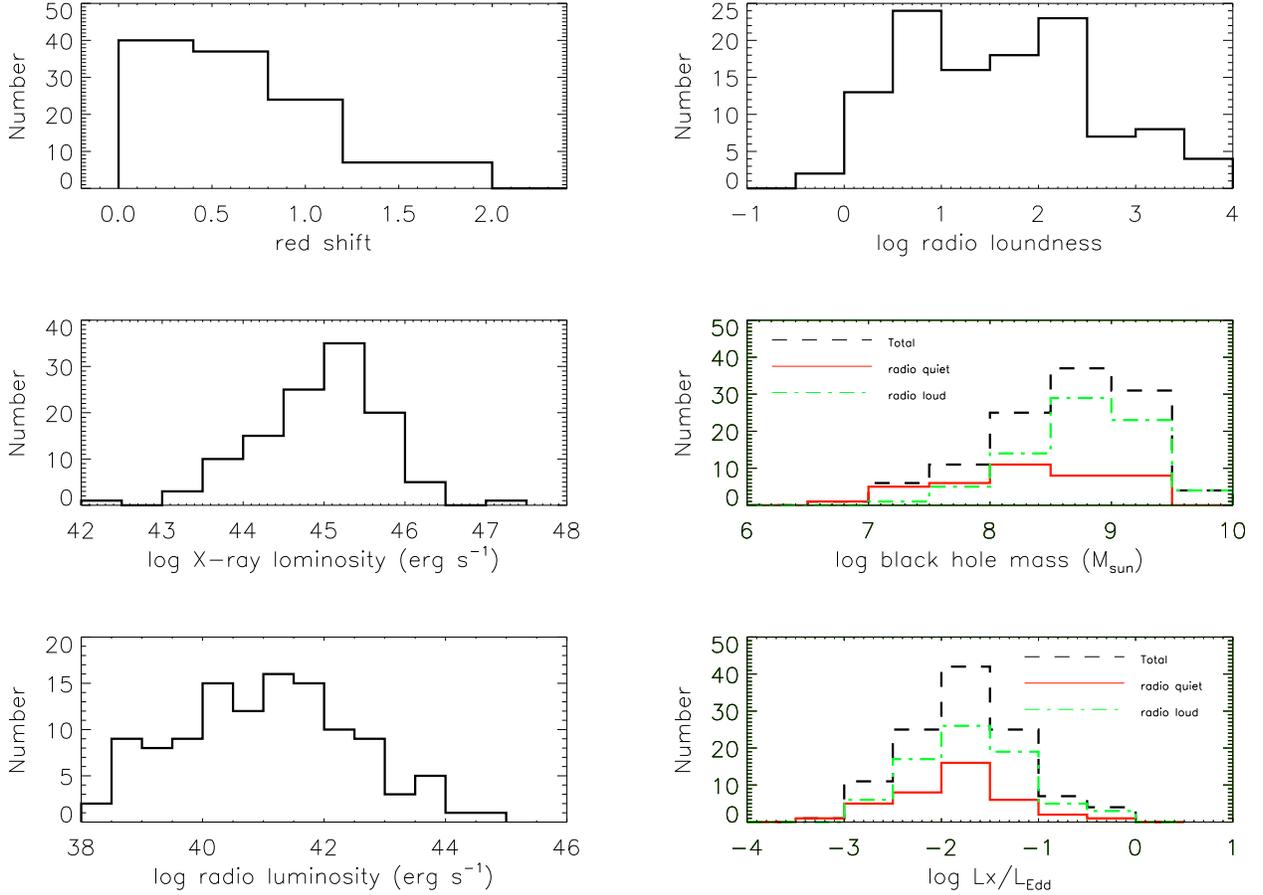} \caption{The global properties of the AGN
sample with the upper panel show the histograms of redshift and
logarithm radio loudnesses, middle panel the logarithm X-ray
0.1-2.4keV and black hole mass (in unit of $M_{\odot}$), and
bottom panel the rest frame 1.4GHz radio luminosities and ratios
of X-ray to Eddington luminosity
($L_{Edd}=1.26\times 10^{38}(M/M_{\odot})
erg\,s^{-1}$). For the mass and luminosity ratio histograms, the
dashed lines give the distributions of the total sample while the
solid and  dash-dotted lines represent the radio quiet and
radio loud sub-sample. \label{six}}
\end{figure}

In this part we list some properties of our broad line AGN sample.
Table~\ref{tbl-1} gives the total 115 sources with the SDSS optical
source name, redshift, logarithm black hole mass in M$_\odot$,
logarithm 0.1-2.4keV X-ray luminosity, logarithm rest frame 1.4GHz
radio luminosity, and logarithm radio loudness in different columns.
The sample include 39 radio quiet and 76 radio loud AGNs. In
Fig.~\ref{six} we show the histograms of redshift, logarithm radio
loudness (LogR), 0.1-2.4keV X-ray (L$_X$) and 1.4GHz radio (L$_r$)
luminosities, black hole mass (M), and the logarithm ratio of X-ray
to the Eddington luminosity. The redshift range for our sample is
from 0.04 to nearly 2. The logarithm black hole masses distribute
from about 6.7 to 9.7. Both the X-ray and radio luminosity
distributions are in a broad range and cover more than 5 orders of
magnitude. Additionally, the X-ray to Eddington luminosity ratio
distributes from about 10$^{-3.5}$ to 1.

\begin{figure}
\centering \plottwo{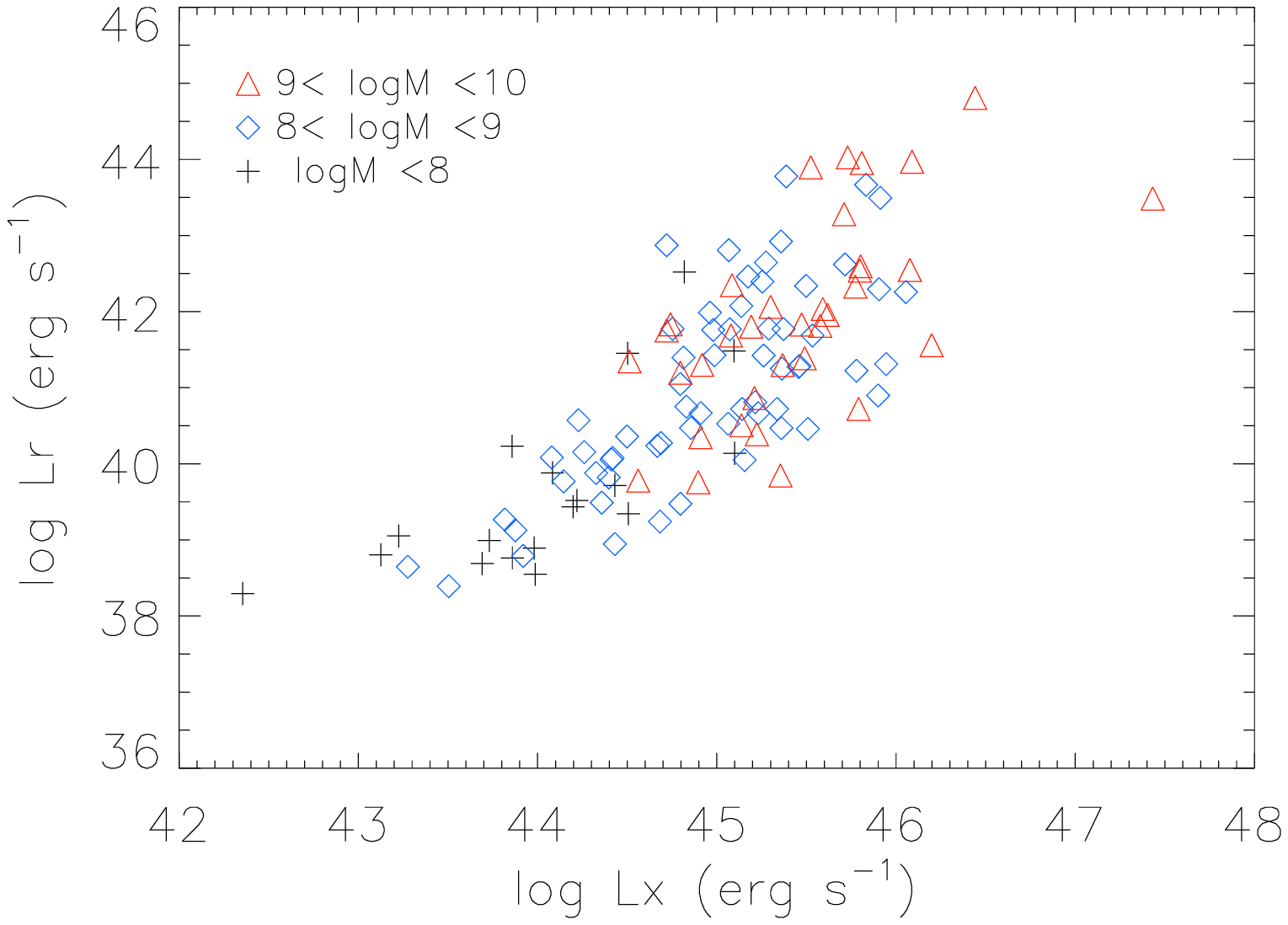}{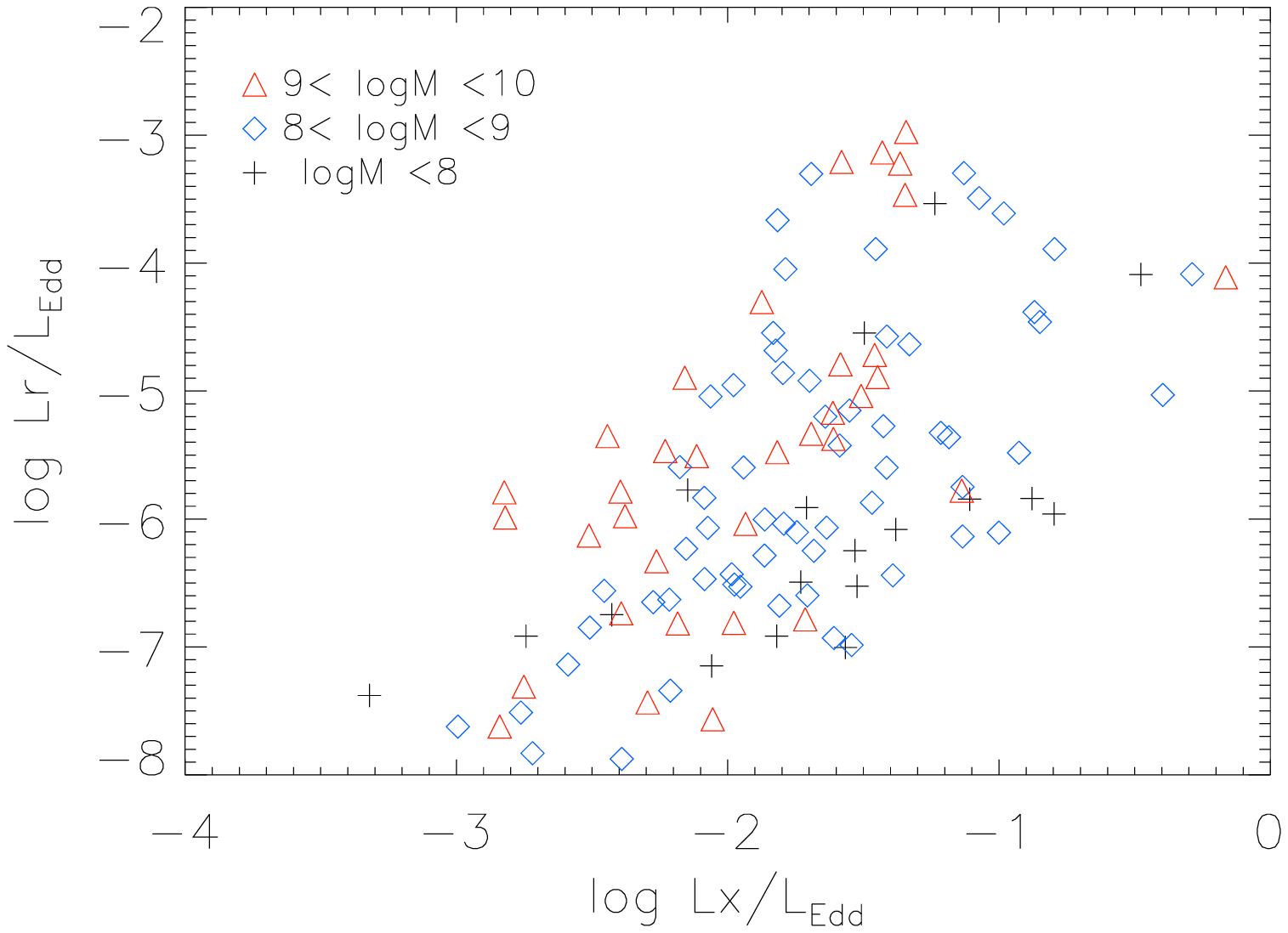} \caption{The 1.4GHz rest
frame radio luminosity versus the 0.1-2.4keV X-ray luminosity with
different symbols corresponding to different bins of logarithm black
hole mass (in unit of $M_{\odot}$). In the left panel we plot the 
logarithm luminosities
directly while in the right panel we scale the luminosity with the
Eddington luminosity. \label{mbin}}
\end{figure}

In Fig.~\ref{mbin} we plot the radio luminosity against the X-ray
luminosity with different
symbols denoting different black hole mass bins. In the left
panel we plot the logarithm luminosity while in the right panel
we scale the luminosity with the Eddington luminosity. We do not see
the clear trends that tracks of different mass bins are parallel
to each other as that found in the former study \citep{mhd03}.

\begin{figure}
\centering \plottwo{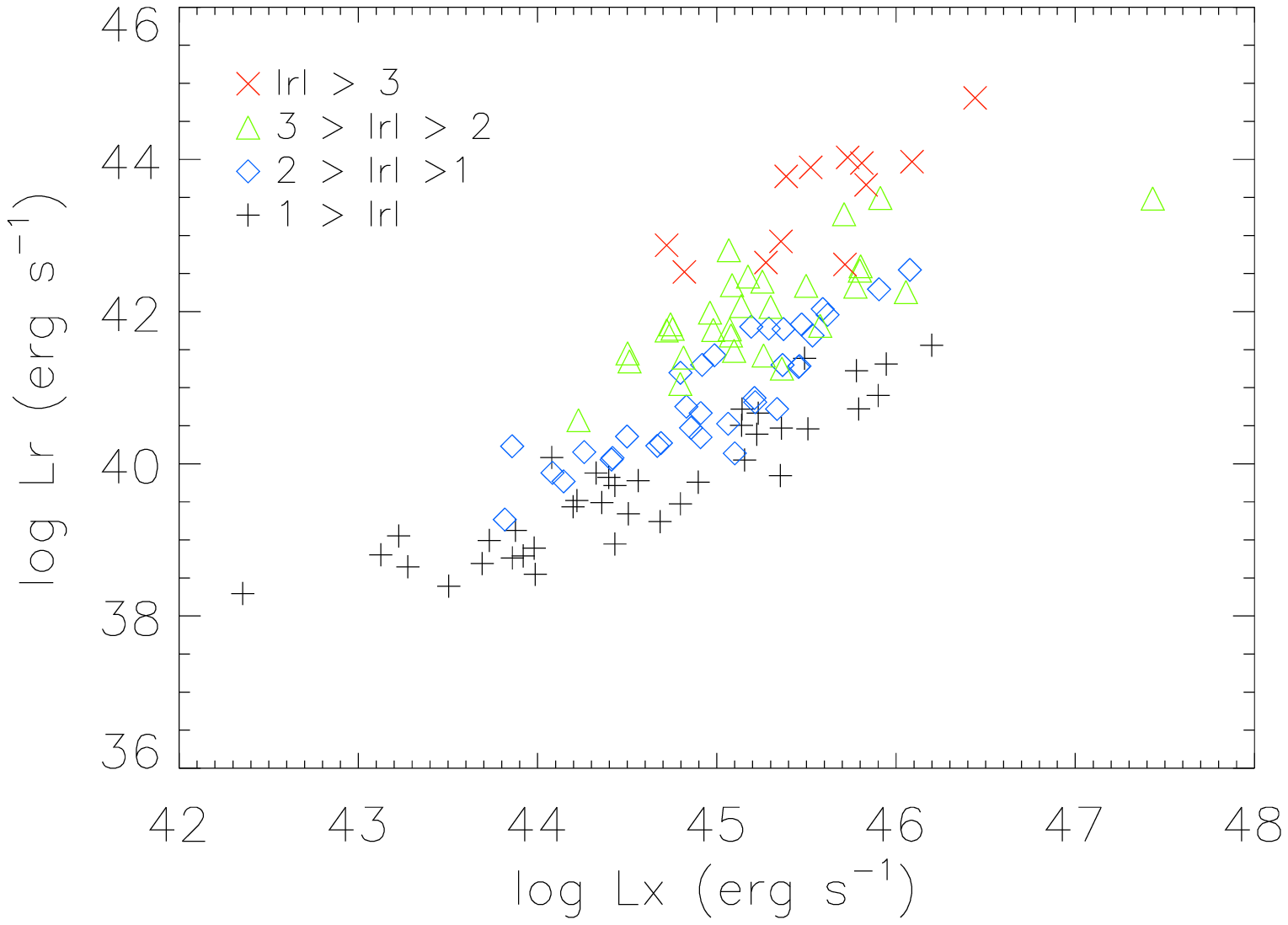}{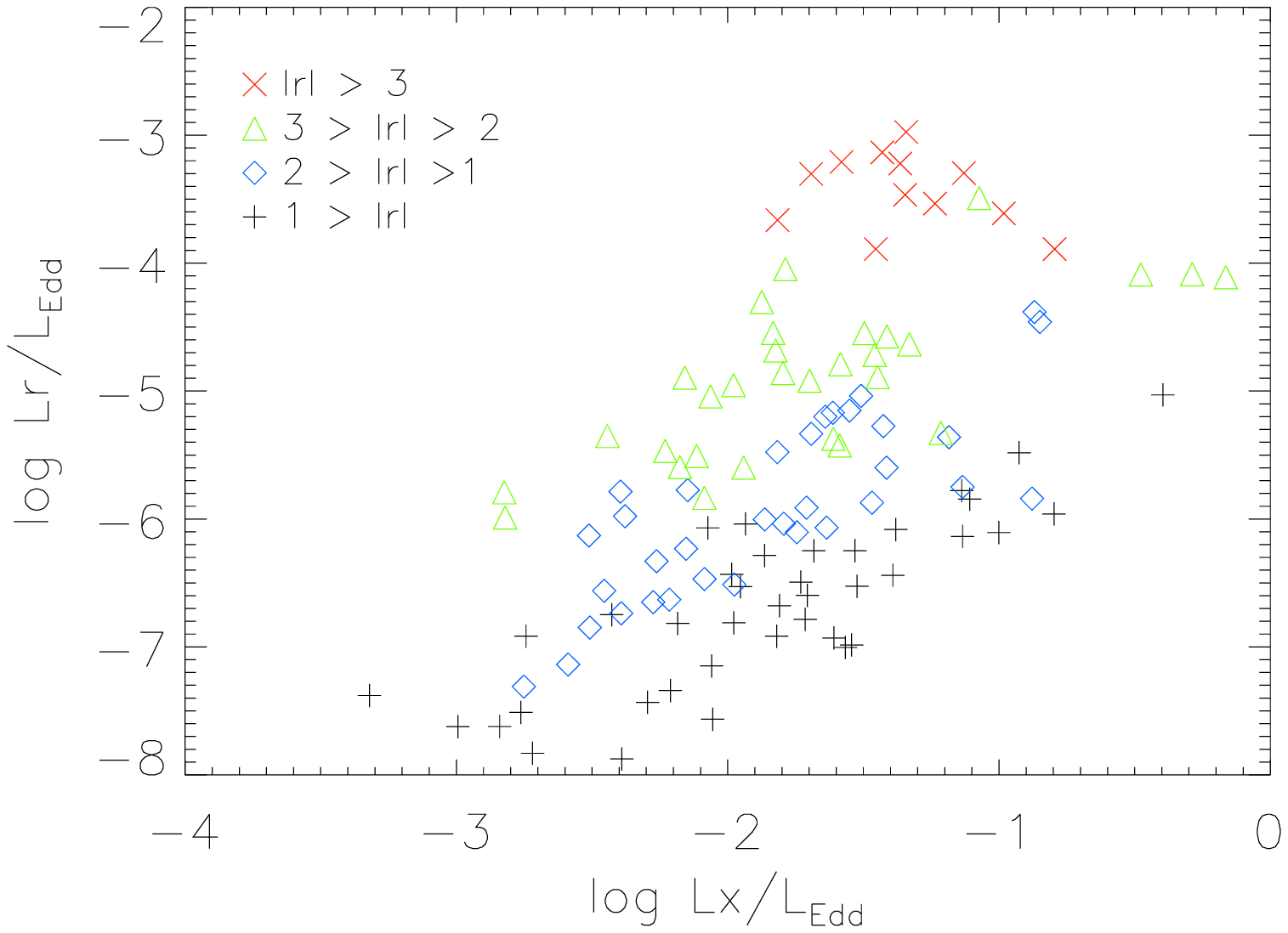} \caption{The rest frame 1.4GHz
radio luminosity versus 0.1-2.4keV X-ray luminosity with
different symbols represent different bins of logarithm radio
loudness (lrl) . In the left panel we plot the logarithm
luminosities directly while in the right panel we scale the
luminosity with the Eddington luminosity. \label{rbin}}
\end{figure}

In Fig.~\ref{rbin} we re-plot the radio and X-ray luminosity,
according to different radio loudness bins. We find that the
sources obviously distributes in different parallel tracks
according to different radio loudness.

\section{Statistic results}

In this section we present the correlation tests for the three
fundamental plane variables, L$_r$, L$_X$, and M, and then give
the fitting results.

\subsection{Correlation tests}

First of all, we perform the partial Kendall's $\tau$ test
\citep{as96} as  did in \citet{mhd03} to test the intrinsic
correlation among the three fundamental plane parameters (i.e. M,
L$_X$, and L$_r$ ), and in addition, among the Eddington luminosity
scaled luminosities and black hole mass. Table~\ref{tbl-2} shows the
results of this test, namely whether the correlation between X and Y
is intrinsic or is only introduced by a third variable Z. The first
three columns list the variables, the forth and fifth columns give
the sub-sample type and number of object, and the last three columns
list the partial Kendall's $\tau$ correlation coefficient, the
square root of the calculated variance $\sigma$, and the probability
to accept the null hypothesis. The null hypothesis will be rejected
with a probability less than the significance level (i.e.
$\sim0.05$).

\begin{figure}
\centering \plotone{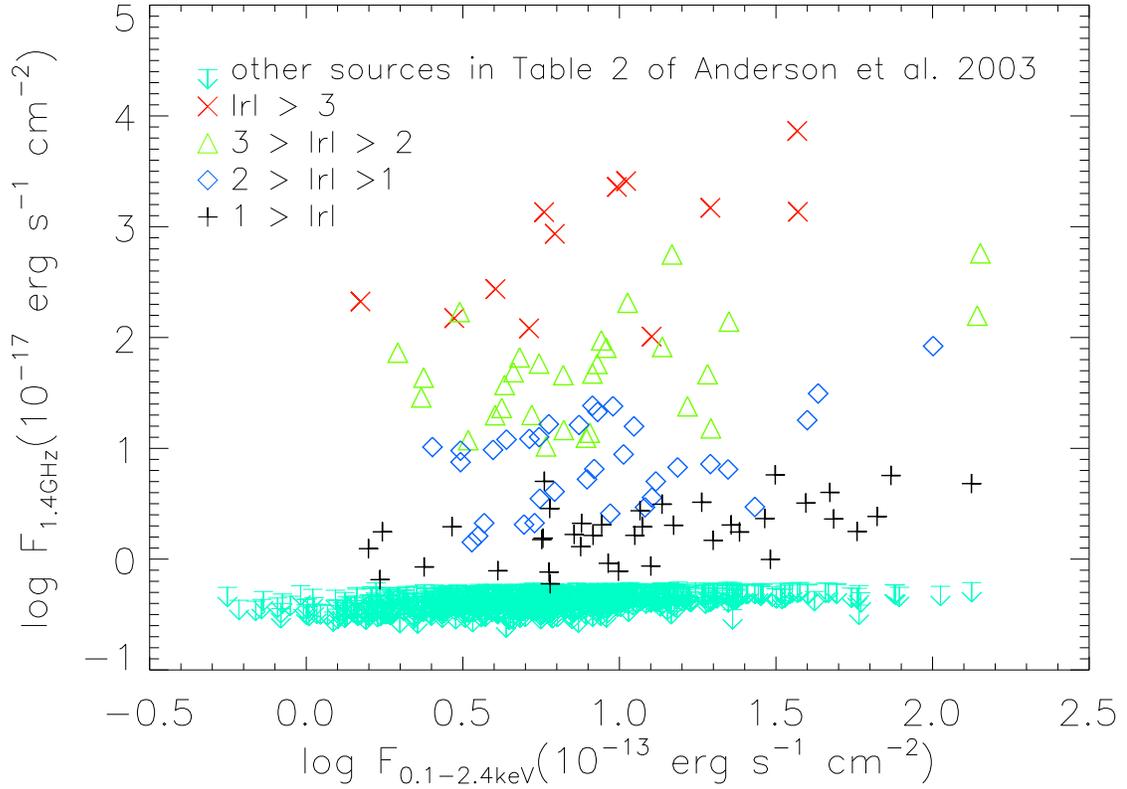} \caption{The rest frame 1.4GHz radio
flux versus 0.1-2.4keV X-ray flux for the broad permitted line AGN
sample in Table 2 of \citet{a03}. For the 115 sources in our sample,
different symbols represent those in different bins of logarithm radio
loudness. For other sources without FIRST detection, we take 0.45mJy as
an upper limit of the observed 1.4GHz radio flux
\citep{becker03}.\label{frx}}
\end{figure}

For the correlation between luminosities, the same dependence on the
source distance always confuses the intrinsic physical relation.
This is the main reason why the reality of the black hole
fundamental plane relation is suspected \citep{b05}. However, the
partial correlation tests indeed prove that the luminosities are
still strongly correlated even if the effect of distance is
excluded.  Furthermore, in order to avoid the distance effect
we  test the existence of the
intrinsic relationship between radio and X-ray emissions by
comparing the radio and X-ray fluxes of the AGNs in our sample. 
In Fig.~\ref{frx} we plot the  rest frame
1.4GHz  radio flux versus the 0.1-2.4keV X-ray flux. We also plot 
the data of all the other sources (without FIRST detection) 
of the broad
permitted line AGN sample in Table 2 of \citet{a03} as a comparison.
For these sources we take 0.45mJy, the typical 3 $\sigma$ detection
level of the FIRST 20cm survey \citep{white97, becker03}, as an
upper limit of radio emission. We see that for these upper limit sources no
correlations exist in the flux-flux plot, while for sources in our sample 
the radio and X-ray fluxes are clearly correlated in each radio loudness
bins. For radio-quiet sources and radio-loud sources in 3 different 
radio-loudness bins, the Spearman correlation coefficients are 0.56, 0.27,
0.37 and 0.57 respectively. The corresponding power-law slope values for these
different samples, derived from 
the Ordinary Least Square (OLS) bisector method \citep{i90}, are 
$0.60\pm0.07$, $1.10\pm0.09$,  $1.05\pm0.16$ and  $1.40\pm0.30$.
Therefore,  the intrinsic correlation between the radio and 
X-ray emissions of AGNs really exist, though the distance effect can stretch 
out the luminosity relationship significantly and affect the intrinsic 
relation seriously in a flux-limited AGN sample.

In addition, from Table 2 we can see that the correlations between the 
luminosities (L$_r$ or
L$_X$) and black hole masses still exist in the partial correlation
test when taking the other luminosity (L$_X$ or L$_r$) as the third
variable. However, when considering the luminosities scaled with the
Eddington luminosity, we see that for the radio quiet sample the
correlation with the black hole mass disappears.

\subsection{The black hole fundamental plane}

Based on the correlation analysis above, we finally fit the data
in the form of:
\begin{equation}
Log(\frac{L_r}{10^{40}erg\,s^{-1}})=\xi_{RX}Log(\frac{L_X}{10^{44}erg\,s^{-1}})+\xi_{RM}Log(\frac{M}{10^8M_{\odot}})+Const.
\end{equation}
We also perform the fitting directly between the Eddington luminosity
scaled radio and X-ray luminosities for the radio quiet sample in
the form of:
\begin{equation}
Log(\frac{L_r}{L_{Edd}})=\xi_{ERX}Log(\frac{L_X}{L_{Edd}})+Const.
\end{equation}

We apply the OLS multivariate regression
method  to the total sample and to the radio quiet and
radio loud subsamples respectively. Table~\ref{tbl-3} summarizes our
OLS bisector fitting results and lists the fitting parameters in equation (1)
with errors in one-sigma confidence level  and the dispersions
($\sigma_r$)\footnote{We define the dispersion as the square root of
the variance of the differences between the observed radio
luminosity and that calculated from the fitting relation. }. The
fitting result from Merloni et al. (2003) is also listed for
comparison.  Compared with the fundamental plane relation of
\citet{mhd03}, our results from the total sample of 115 sources and
both the radio loud and radio quiet subsamples give a much smaller
black hole mass dependence ($\xi_{RM}$). And much larger
radio--X-ray correlation slopes ($\xi_{RX}>1$) are found in the
fittings of the total sample and the radio loud subsample when
compared with the value of $\xi_{RX}=0.60$ in \citet{mhd03}. The
radio--X-ray correlation slope ($\xi_{RX}=0.85$) in the result of
radio quiet subsample is also  larger than that in
\citet{mhd03}.

The result of the Eddington luminosity scaled
luminosity fitting for radio quiet sources is:
\begin{equation}
Log(\frac{L_r}{L_{Edd}})=(0.86\pm0.10)Log(\frac{L_X}{L_{Edd}})+(-5.08\pm0.19),
\end{equation}
where errors are in one-sigma confidence level.

\begin{figure}
\centering \plottwo{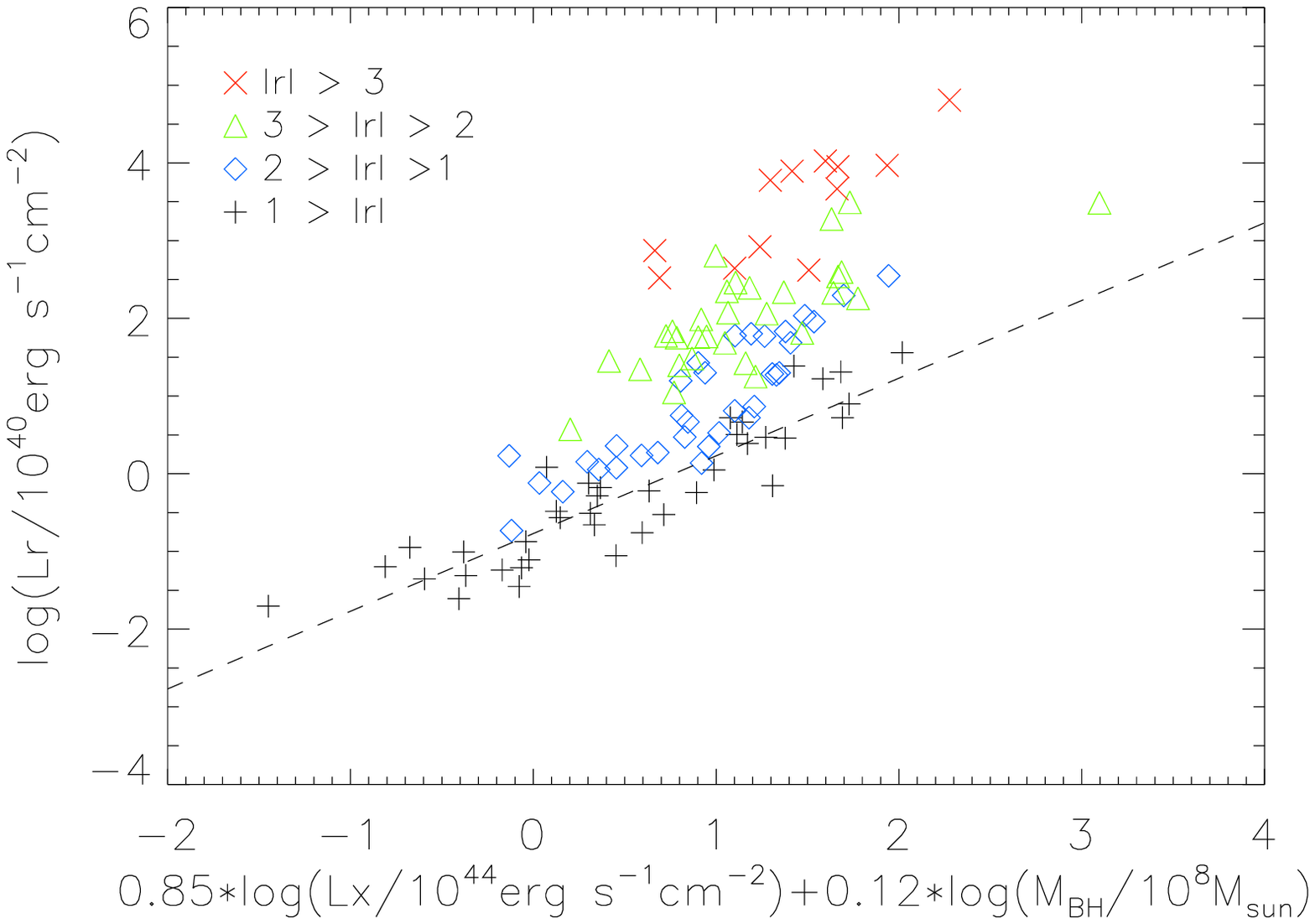}{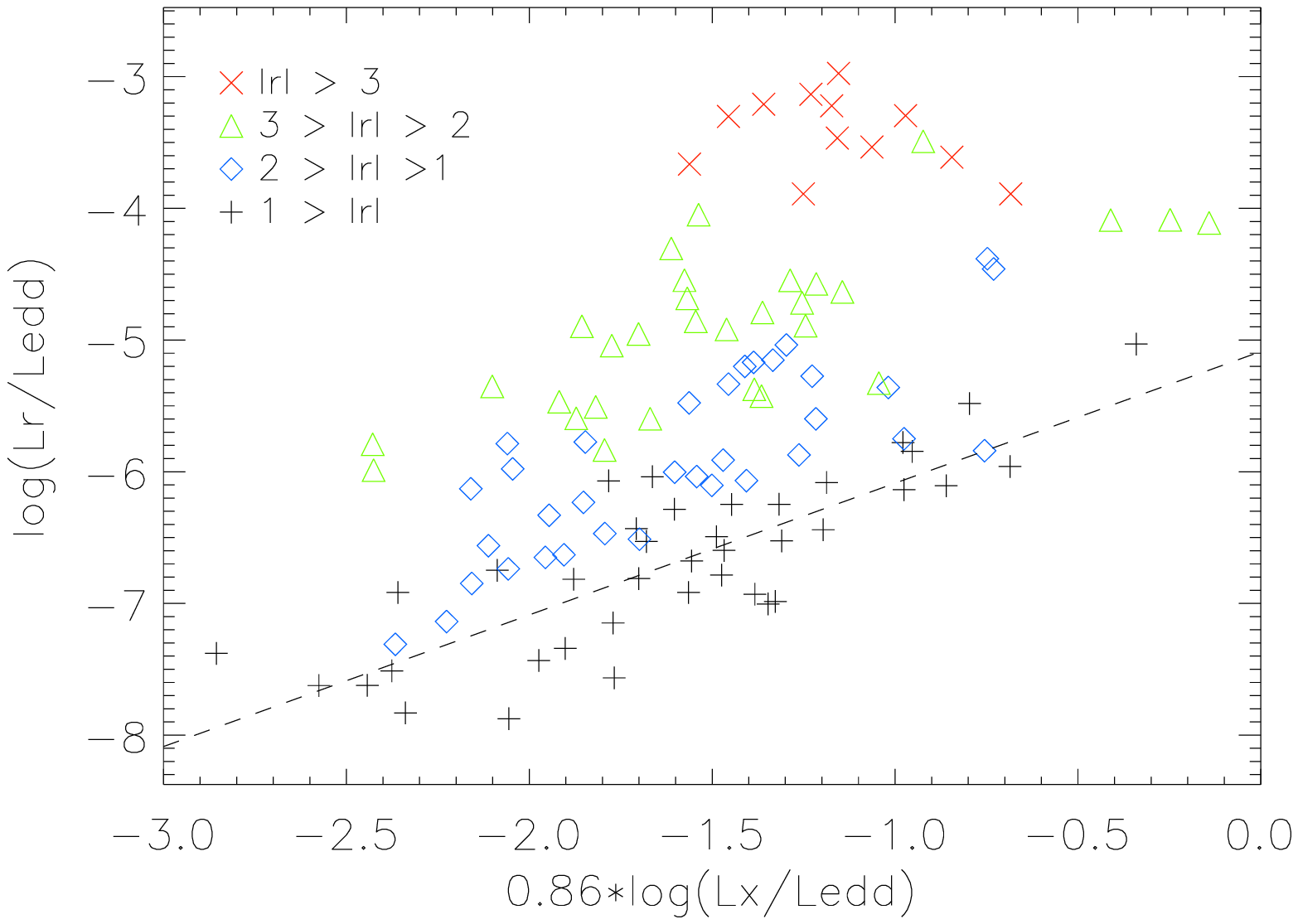} \caption{Fitting results.
The left panel shows all the AGNs in the sample according to the
radio quiet fundamental plane relation with different symbols
corresponding to different logarithm radio loudness (lrl) bins. The
right panel shows the Eddington luminosity scaled radio and X-ray
luminosities according to the fitting result for radio quiet
sources. The dashed line represents the best fitting relation with
the radio quiet sources only. \label{res}}
\end{figure}

Fig.~\ref{res} plots the fitting relations for radio quiet sources
(eq. (4) and (6)) plus all the data points (including radio-loud sources)
in different radio
loudness bins with the x coordinate represents the quantity calculated
from  these two equations for radio-quiet sources.

\begin{figure}
\centering \plotone{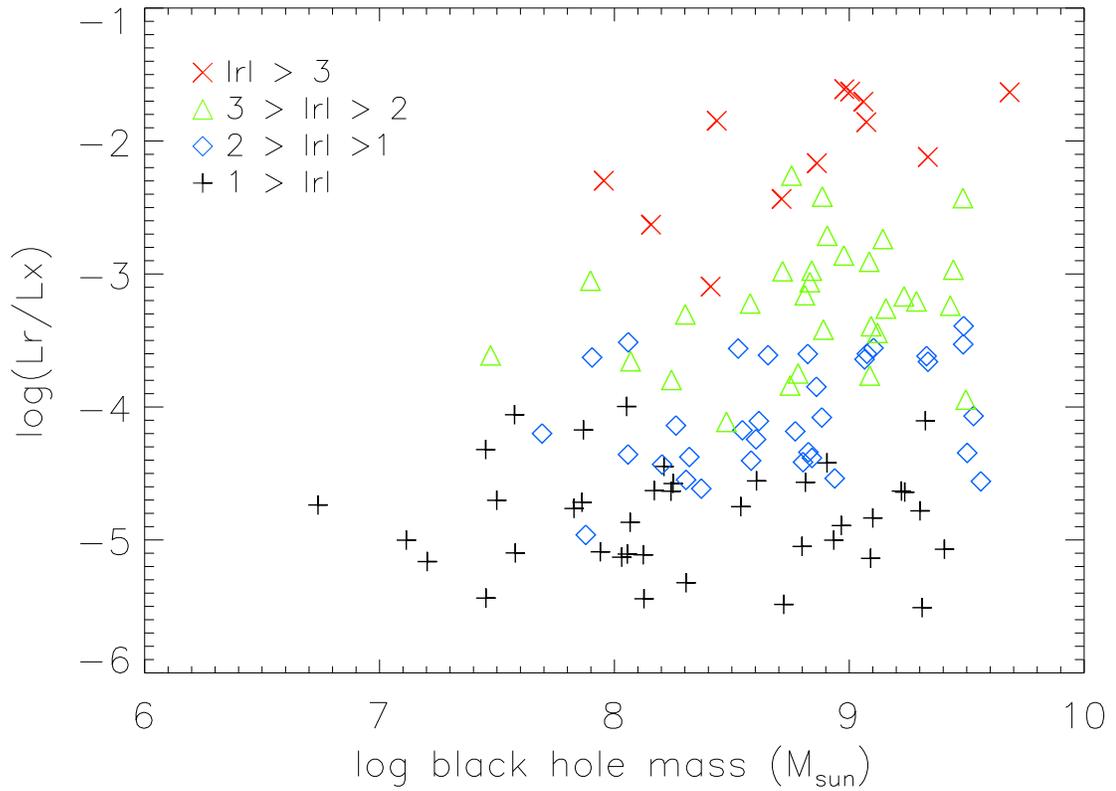}\caption{The ratios of radio and X-ray
luminosity are ploted versus the black hole masses in logarithm. 
Different symbols represent sources in different bins of logarithm radio
loudness.\label{Lrxm}}
\end{figure}

The first thing we find out with our data is that our fundamental
relationships are not sensitive to the black hole mass no matter the
sources are radio loud or radio quiet (see Table~\ref{tbl-3}). The
correlation slope value $\xi_{RM}$ is nearly zero within
uncertainties for all the three fittings, which is different from
what was found by \citet{mhd03}. This is not surprise as our sample
is uniform and includes only supermassive black holes whose masses
are in a much narrower range compared to those in \citet{mhd03}. In
this case the black hole mass may not affect the relation
significantly.  In Fig.~\ref{Lrxm} we plot $L_r/L_X$ ratio
versus black hole mass (M) in logarithm, which shows that the
dependence of radio to X-ray luminosity ratio on black hole mass is
also very weak.

Secondly we get very different radio--X-ray correlations for the
radio loud and radio quiet samples. The slope $\xi_{RX}$ for our
radio quiet sample is about 0.85 which is a little steeper than the
result in \citet{mhd03}, but still close to the value found in
the hard state of GBHs (Gallo et al. 2003, $\xi_{RX}\sim 0.7$). For the
total sample and the radio loud subsample, the parameter
$\xi_{RX}$ is around 1.33 and 1.39 respectively. Thus either a much
different radiation mechanism or some additional effects must be
taken into account for
the radio loud sources.

In addition, we also obtain a tight relation between the Eddington
luminosity scaled radio and X-ray luminosities for the radio quiet
subsample with the correlation slope $\xi_{ERX}=0.86$ and a
dispersion $\sigma_r = 0.38$.

\section{Discussions}

Previous studies on the black hole fundamental plane relation mainly
concerned about the relations of black hole mass, hard X-ray
luminosity (eg. 2-10keV), and radio luminosity at or above 5GHz
\citep{gfp03, mhd03, yc05}. In this work, we use the 0.1-2.4keV
X-ray luminosity and 1.4GHz radio luminosity instead. If we assume
that the X-ray and radio emission can be described as power laws
with the typical spectral index  in each band for all the sources,
the corrections from hard to soft X-ray and from 5GHz to 1.4GHz
should be linear and only change the constant item in the
fundamental plan relation (Eq. (4)). However, the actual spectrum of
individual sources in our sample is probably more complicated,
especially in the soft X-ray band. The spectral parameters may
change from one source to another according to different luminosity
level or different dominated radiation mechanisms at different
accretion rate. Thus any nonlinear relations between the hard/soft
X-ray luminosities or 5GHz/1.4GHz radio luminosities for sources in
our sample will change the correlation slopes and result in
different correlations in our fittings from those obtained in
previous works. On the other hand, the small differences in the
radio and X-ray spectral index for radio-quiet and radio-loud
subsamples \citep{y98} may also lead to some differences in the
derived black hole fundamental plane. Future detailed studies on the
radio and X-ray spectral index for individual sources in our sample
are expected to resolve this problem.

In this work we found very different black hole
fundamental plane relations for radio loud and radio quiet broad
line AGNs. Although whether such a difference is intrinsic or not
deserves further studies, our results on the fundamental plane
relation is probably helpful to investigate the underlying black hole
accretion physics and may be used to constrain the  theoretical models.

The fundamental plane relation with the X-ray emission dominated by
the relativistic jet has been well discussed in literature.
\citet{h04} recently showed that the synchrotron emission from the scale
invariant jet including the effect of radiation cooling can lead to
a correlation of $L_{R}\propto M^{0}L_{X}^{1.42}$ with typical scale
invariant synchrotron jet parameters. Similar conclusions were also
obtained from studies on Galactic black hole X-ray binaries \citep{y05,
yc05}. Our derived fundamental plane correlation slope for radio
loud sample seems consistent with this jet dominate X-ray model as
well.

However, for radio loud AGNs the Doppler beaming effect from the
relativistic jet should also be important.  Doppler beaming can
increase the jet intrinsic power by a factor of
$\delta^{2+\alpha}$ (where $\delta$ is the beaming factor and
$\alpha$ is the intrinsic power-law spectra index, i.e.
$f_{\nu}\propto \nu^{-\alpha}$, \citet{k99}) and may lead to a
spurious observed correlation. There are two situations needed to
be discussed when thinking about the beaming effect. One is that
both the radio and X-ray fluxes are dominated by jet emission and
they will be affected by the relativistic beaming to different
levels according to the spectral index in different band. The
other is that only the radio emission is from the jet and beamed
while the X-ray emission is dominated by the accretion flow.
\citet{fkm04} have studied the low-power accretion black hole
systems where the radio and X-ray emissions are both jet
dominated. They used a sample including both the stellar black
hole systems and AGNs, and found that the scaling relation between
radio and X-ray emissions is only slightly dependent on the beaming factors
in the form of shifting along the fitted correlation line.
However, such a jet dominated circumstance only exists in the low
state of black hole systems \citep{gfp03, yc05} and  is thus
possibly not the case for our sample. In our case for broad line
AGNs, probably only the radio emission is beamed. Therefore,
for AGNs with different radio-loudness this extra
effect of relativistic beaming will cause parallel shifts from the
intrinsic fundamental plane, and thus lead to a steeper correlation slope in
the observed correlation.

\begin{figure}
\centering \plottwo{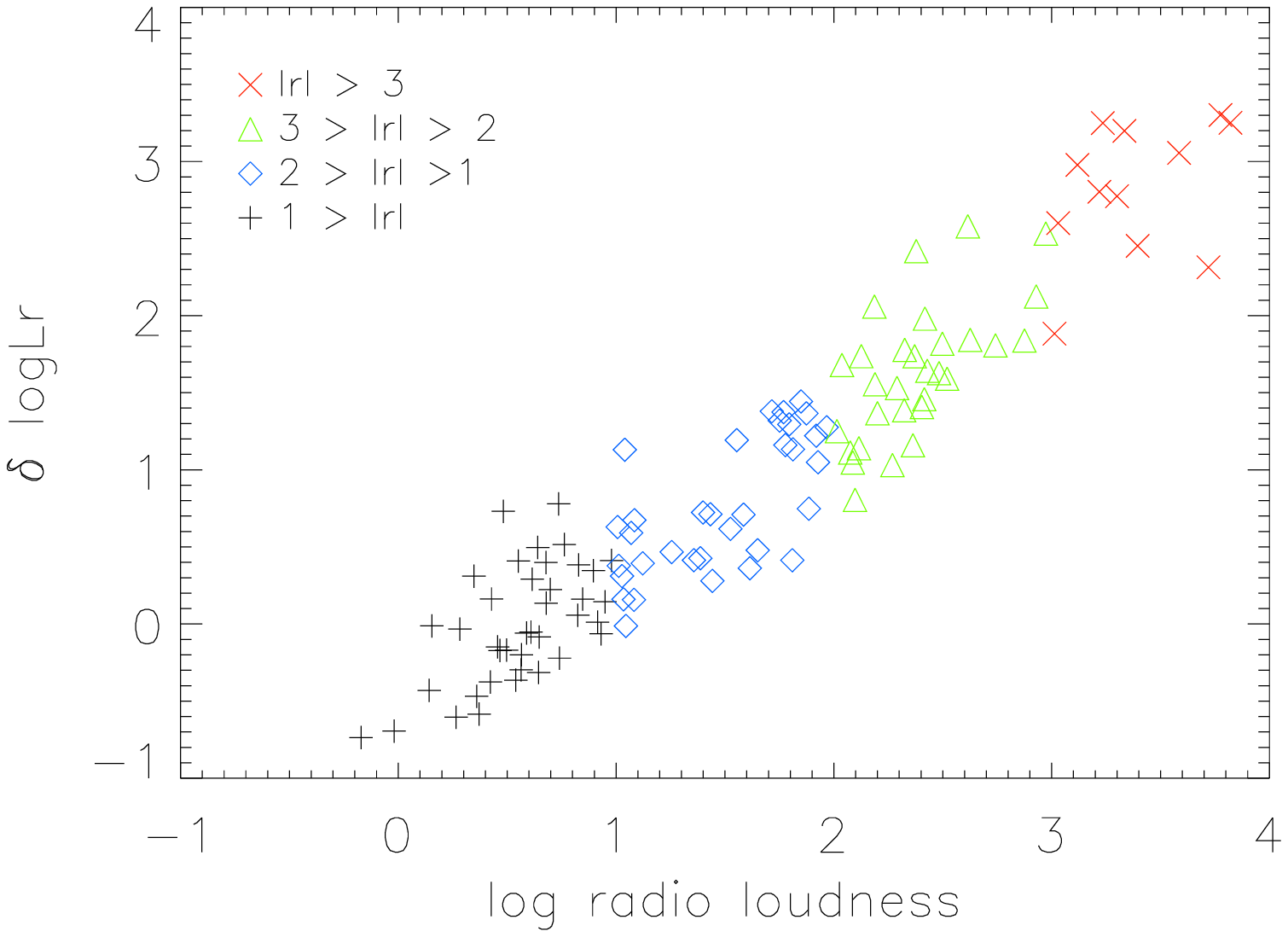}{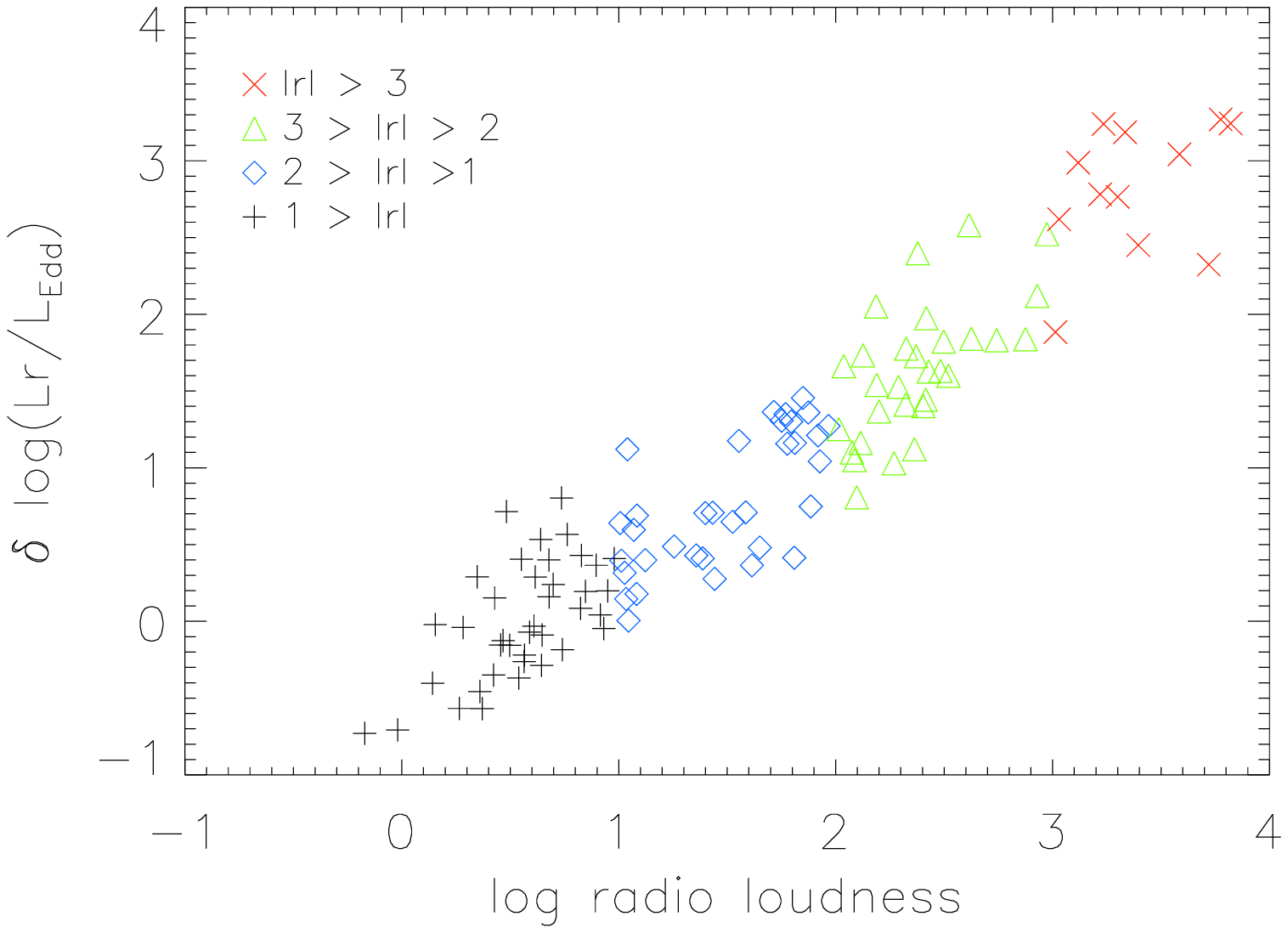} \caption{The differences
($\delta Log L_r$) between the observed radio luminosity and that
derived from the fundamental plane of the radio quiet AGNs for all
AGNs in the sample with different logarithm radio loudness (lrl)
bins are ploted versus the logarithm radio loudness. The left panel
gives the logarithm radio luminosity difference calculated with
equation (4) using the fitting result from the radio quiet sources
in Table~\ref{tbl-3}, while the right panel gives the differences
calculated with equation (6). \label{detr}}
\end{figure}

We noticed that the beaming factor is hard to be
measured directly for large sample of AGNs, so we use the radio
loudness as a possible indicator of it. In Fig.~\ref{rbin} and
Fig.~\ref{res} all radio loud sources are shown in a parallel
distribution according to different radio loudness bins, which
indicates the possibility that all of these sources may obey the
same correlation as that of the radio quiet sources but it is
shifted by a different level according to different radio loudness.
If we take the radio quiet fundamental plane relation as an
intrinsic one for all AGNs with high accretion rate, the difference
($\delta Log L_r$ or $\delta Log (L_r/L_{Edd})$) between the
observed radio luminosity and that derived from the radio quiet
fundamental plane can be calculated. In Fig.~\ref{detr} we plot
$\delta Log L_r$ and $\delta Log(L_r/L_{Edd})$ versus radio
loudness and tight correlations are found. As shown in this plot,
the differences between the observed and estimated radio luminosity
can be different in more than 3 orders of magnitude. This is not
impossible for the extremely radio loud sources with a beaming
factor greater than 10.  Also the difference between the observed and
calculated radio luminosities are tightly correlated with the radio
loudness. Table~\ref{table4} lists the partial Kendall's $\tau$
correlation coefficient between $\delta Log L_r$, $\delta
Log (L_r/L_{Edd})$ and logarithm radio loudness when taking the
logarithm observed radio luminosity as the third variable. The test
shows that the correlation  still exists when the effects of
radio luminosity are excluded. Therefore we think that the observed
fundamental plane for radio loud AGNs is unreliable unless the
beaming effect can be removed.

When considering the fundamental plane for radio quiet sources only,
the underlying physical mechanism is not very clear as well. We
compared our results with the available accretion flow models
discussed in \citet{mhd03} with corresponding parameters listed in
Table 3 of their paper. As the ROSAT observations are in the soft
X-ray band, we also investigated the model of multicolour thermal
emission from the inner part of a standard thin disk \citep{ss73}.
Moreover, the jet dominated situation with radiation cooling effect
is also considered. At last we found that our fitting result can be
marginally matched when the magnetic field dependence on black hole
mass and accretion rate in the form of $B^2\propto M^{-1}\dot{m}$
and the X-ray luminosity has a non-linear dependence on accretion
rate with a power-law index around 2. Such parameters can be
satisfied in the radiatively inefficient accretion flow models
\citep{n98}. Similar conclusion was also found in \citet{mhd03}.
However, there is no evidence that the sources with radiatively
inefficient accretion flow occupy a large percentage in our sample
when considering the X-ray to Eddington luminosity ratio range
($10^{-3}$ to 1, see the $L_X/L_{Edd}$ histogram in Fig.~\ref{six}).
We noticed that the emission in soft X-ray band is complex and may
come from different radiation mechanisms \citep{bs97} including the
multicolor thermal emission from the inner part of an optically thick
disk, inverse Compton emission, and free-free emission in the hot
corona. So more analyses on the X-ray and radio spectra as well as
accretion disk model studies are needed for obtaining more accurate
physical explanations of such a fundamental plane relation.

\section{Summary}

With a uniform broad line AGN sample we have studied the fundamental
plane relation in black hole accretion systems. Compared to previous
works, we found that our fundamental plane relation has a much
weaker dependence on the black hole mass. A very tight relation is
also found between the Eddington luminosity scaled X-ray and radio
luminosities for radio quiet sources. The non-linear dependence of
radio luminosity on X-ray luminosity is different for radio loud and
radio quiet AGNs. We attribute this to the relativistic jet in
radio-loud AGNs. Especially the Doppler beaming effect from a
relativistic jet can increase the intrinsic radio emission
significantly and lead to a steeper slope of the fundamental plan
relation. The derived fundamental plane relation for radio-quiet
sources should better reflect the intrinsic physical correlations.
However, the soft X-ray emission mechanisms of AGNs are complex and
more reliable theoretical explanations are expected to be obtained
only after further detailed spectral studies in soft X-ray and radio
band on the individual AGNs in our sample can be done.  Finally, we would
like to mention that our black hole fundamental plane relation is
derived from a flux-limited sample consisting of 115 broad line AGNs.
Obviously more future studies with larger and better samples are needed
to confirm our result.

\acknowledgments
We thank the anonymous referee for valuable suggestions and Fukun Liu, 
Lei Qian, Bingxiao Xu and 
Feng Yuan for helpful discussions. This work
is supported by the NSFC funds (No. 10473001 \& No. 10525313), the RFDP
Grant (No. 20050001026) and the Key Grant Project of Chinese Ministry
of Education (No. 305001).

\clearpage


\begin{deluxetable}{lccccr}
\tabletypesize{\scriptsize} \tablewidth{0pt} \tablecaption{The AGN
sample \label{tbl-1}}

\tablehead{ \colhead{Name}           & \colhead{z} &
\colhead{Log($\frac{M}{M_{\odot}}$)}          &
\colhead{Log($\frac{L_X}{ergs^{-1}}$)} &
\colhead{Log($\frac{L_r}{ergs^{-1}}$)} & \colhead{LogR}}
\startdata
SDSS J000608.04-010700.8 &     0.949  &     8.770  &    45.456  &    41.272  &     1.433 \\
SDSS J000710.01+005329.0 &     0.316  &     9.091  &    44.896  &    39.758  &     0.539 \\
SDSS J000813.22-005753.3 &     0.139  &     8.305  &    43.817  &    39.268  &     1.082 \\
SDSS J003847.96+003457.4 &     0.081  &     8.170  &    43.276  &    38.647  &     0.915 \\
SDSS J004319.74+005115.3 &     0.308  &     9.301  &    44.561  &    39.779  &     0.647 \\
SDSS J005441.19+000110.6 &     0.647  &     8.370  &    45.336  &    40.721  &     1.027 \\
SDSS J005905.51+000651.5 &     0.719  &     8.979  &    45.387  &    43.777  &     3.821 \\
SDSS J012100.73-001519.0 &     0.864  &     8.156  &    45.274  &    42.645  &     3.720 \\
SDSS J012240.12-003239.7 &     0.883  &     8.966  &    45.360  &    40.469  &     0.283 \\
SDSS J012528.83-000555.9 &     1.077  &     9.335  &    46.089  &    43.969  &     3.219 \\
SDSS J012905.32-005450.6 &     0.707  &     8.527  &    44.987  &    41.427  &     1.795 \\
SDSS J014644.82-004043.2 &     0.083  &     6.738  &    43.730  &    38.993  &     0.949 \\
SDSS J015105.80-003426.3 &     0.335  &     8.803  &    44.688  &    40.273  &     1.614 \\
SDSS J015950.24+002340.9 &     0.163  &     8.262  &    44.499  &    40.359  &     1.084 \\
SDSS J020615.99-001729.1 &     0.043  &     8.123  &    43.504  &    38.392  &     0.142 \\
SDSS J021225.57+010056.1 &     0.513  &     8.716  &    44.753  &    41.774  &     2.498 \\
SDSS J022347.48-083655.5 &     0.261  &     7.828  &    44.198  &    39.435  &     0.824 \\
SDSS J024240.31+005727.2 &     0.569  &     9.405  &    45.791  &    40.721  &     0.565 \\
SDSS J073623.12+392617.8 &     0.118  &     8.126  &    44.683  &    39.241  &     0.371 \\
SDSS J074242.18+374402.0 &     0.806  &     8.860  &    45.534  &    41.686  &     1.927 \\
SDSS J075047.32+413033.5 &     1.184  &     8.605  &    45.779  &    41.223  &     0.551 \\
SDSS J075407.95+431610.5 &     0.348  &     9.502  &    45.210  &    40.865  &     1.387 \\
SDSS J075819.68+421935.1 &     0.211  &     8.305  &    44.797  &    39.475  &     0.360 \\
SDSS J075838.14+414512.4 &     0.094  &     7.452  &    43.125  &    38.804  &     0.828 \\
SDSS J080131.96+473616.0 &     0.157  &     8.883  &    44.830  &    40.751  &     1.585 \\
SDSS J080322.48+433307.1 &     0.276  &     8.203  &    44.668  &    40.236  &     1.357 \\
SDSS J083317.46+512422.3 &     0.591  &     8.055  &    45.155  &    40.049  &     0.497 \\
SDSS J083525.17+482656.3 &     1.301  &     8.241  &    45.945  &    41.311  &     0.678 \\
SDSS J084224.91+514501.1 &     0.797  &     8.813  &    45.231  &    40.665  &     0.615 \\
SDSS J085442.00+575730.0 &     1.318  &     9.073  &    45.809  &    43.952  &     3.585 \\
SDSS J085457.22+544820.5 &     0.256  &     7.499  &    44.219  &    39.517  &     0.847 \\
SDSS J090145.28-000051.7 &     1.454  &     8.934  &    45.900  &    40.898  &     0.589 \\
SDSS J090745.29+532421.5 &     0.711  &     9.529  &    45.367  &    41.299  &     1.399 \\
SDSS J090910.08+012135.7 &     1.024  &     8.885  &    45.912  &    43.494  &     2.973 \\
SDSS J090924.69+521632.6 &     0.410  &     9.443  &    44.720  &    41.752  &     2.371 \\
SDSS J091205.16+543141.2 &     0.448  &     8.826  &    44.418  &    40.078  &     1.122 \\
SDSS J091301.01+525928.9 &     1.377  &     9.236  &    46.199  &    41.558  &     0.347 \\
SDSS J091333.65-004250.9 &     0.426  &     9.075  &    44.797  &    41.197  &     1.777 \\
SDSS J091635.45+541426.9 &     0.284  &     8.069  &    44.228  &    40.572  &     2.115 \\
SDSS J092856.27+013246.0 &     0.284  &     7.203  &    44.506  &    39.343  &     0.740 \\
SDSS J092943.41+004127.3 &     0.587  &     8.799  &    45.508  &    40.459  &     0.456 \\
SDSS J093200.08+553347.4 &     0.266  &     8.615  &    44.260  &    40.154  &     1.007 \\
SDSS J093609.13-002639.7 &     0.141  &     7.114  &    43.691  &    38.689  &     0.467 \\
SDSS J094042.92+021557.3 &     0.386  &     8.938  &    45.063  &    40.526  &     1.442 \\
SDSS J100017.67+000523.7 &     0.905  &     8.410  &    45.715  &    42.620  &     3.013 \\
SDSS J101044.51+004331.3 &     0.178  &     8.721  &    44.431  &    38.946  &    -0.171 \\
SDSS J101502.24+023128.1 &     0.218  &     7.905  &    43.858  &    40.231  &     1.812 \\
SDSS J101527.26+625911.5 &     0.350  &     8.319  &    44.145  &    39.769  &     1.012 \\
SDSS J101557.05+010913.6 &     0.780  &     9.286  &    45.802  &    42.596  &     2.037 \\
SDSS J103214.52+635950.2 &     0.556  &     8.890  &    44.814  &    41.396  &     2.200 \\
SDSS J105342.21-001420.1 &     0.676  &     8.840  &    44.962  &    41.987  &     2.875 \\
SDSS J111221.82+003028.5 &     0.523  &     8.782  &    44.796  &    41.048  &     2.087 \\
SDSS J111231.13-002534.2 &     0.544  &     8.477  &    45.363  &    41.251  &     2.097 \\
SDSS J115024.79+015620.3 &     0.706  &     7.955  &    44.819  &    42.521  &     3.030 \\
SDSS J115043.87-002354.0 &     1.976  &     9.683  &    46.441  &    44.808  &     3.775 \\
SDSS J115542.53+021411.0 &     0.873  &     8.977  &    45.254  &    42.394  &     2.418 \\
SDSS J120332.94+022934.6 &     0.077  &     7.453  &    43.987  &    38.549  &     0.265\\
SDSS J121347.52+000129.9 &     0.962  &     8.811  &    45.498  &    42.338  &     2.126 \\
SDSS J123200.01-022404.7 &     1.043  &     9.004  &    45.524  &    43.895  &     3.235 \\
SDSS J125337.35-004809.5 &     0.427  &     7.862  &    44.431  &    39.715  &     0.679 \\
SDSS J125500.48+034043.0 &     0.437  &     8.578  &    44.980  &    41.759  &     2.483 \\
SDSS J125519.69+014412.3 &     0.343  &     9.310  &    45.355  &    39.845  &    -0.019  \\
SDSS J125945.18+031726.1 &     1.528  &     9.156  &    45.798  &    42.539  &     2.429  \\
SDSS J130554.15+014929.8 &     0.733  &     9.430  &    45.300  &    42.062  &     2.189  \\
SDSS J134113.93-005315.1 &     0.237  &     8.058  &    44.413  &    40.055  &     1.255  \\
SDSS J134739.83+622149.5 &     0.804  &     8.582  &    45.214  &    40.810  &     1.650  \\
SDSS J134948.39-010621.8 &     0.600  &     9.100  &    45.223  &    40.389  &     0.154  \\
SDSS J135351.58+015153.8 &     1.608  &     8.862  &    45.833  &    43.666  &     3.300  \\
SDSS J135425.23-001358.0 &     1.512  &     9.485  &    46.077  &    42.548  &     1.769  \\
SDSS J135527.98+015527.4 &     1.732  &     8.654  &    45.905  &    42.294  &     1.874  \\
SDSS J140104.87+004332.7 &     0.665  &     9.222  &    45.138  &    40.506  &     0.428  \\
SDSS J140127.69+025606.1 &     0.265  &     7.692  &    44.083  &    39.881  &     1.525  \\
SDSS J140710.59-004915.2 &     1.510  &     8.243  &    46.055  &    42.257  &     2.014  \\
SDSS J141556.84+052029.5 &     0.126  &     8.031  &    43.920  &    38.790  &     0.423  \\
SDSS J142339.88+043634.8 &     1.649  &     9.120  &    45.773  &    42.329  &     2.415  \\
SDSS J142519.16+035425.8 &     0.792  &     8.302  &    45.072  &    41.768  &     2.520  \\
SDSS J142545.90+002242.7 &     0.326  &     7.898  &    44.502  &    41.452  &     2.742  \\
SDSS J143244.44-005915.1 &     1.027  &     9.335  &    45.618  &    41.958  &     1.554  \\
SDSS J143641.94+022940.5 &     0.772  &     8.823  &    45.372  &    41.771  &     1.967  \\
SDSS J145002.46+001629.3 &     0.957  &     9.087  &    45.577  &    41.813  &     2.075  \\
SDSS J145126.16+032643.3 &     0.479  &     9.560  &    44.909  &    40.350  &     1.034  \\
SDSS J150759.73+041512.2 &     1.701  &     9.496  &    47.432  &    43.486  &     2.363  \\
SDSS J150935.97+574300.5 &     1.705  &     9.324  &    45.490  &    41.386  &     0.482  \\
SDSS J150940.68+571811.8 &     0.817  &     8.832  &    45.136  &    42.074  &     2.325  \\
SDSS J151441.13+555932.7 &     1.190  &     8.905  &    45.141  &    40.720  &     0.979  \\
SDSS J154751.94+025550.8 &     0.098  &     7.940  &    43.981  &    38.892  &     0.644  \\
SDSS J154929.44+023701.1 &     0.414  &     8.712  &    45.358  &    42.922  &     3.395  \\
SDSS J155607.44+552436.2 &     0.434  &     8.052  &    44.079  &    40.083  &     0.736  \\
SDSS J155620.24+521520.0 &     0.227  &     8.067  &    44.358  &    39.490  &     0.609  \\
SDSS J160623.57+540555.8 &     0.876  &     8.754  &    45.067  &    42.807  &     2.615  \\
SDSS J160713.91+483326.2 &     0.125  &     8.538  &    43.876  &    39.126  &     0.931  \\
SDSS J160732.86+484619.9 &     0.146  &     7.868  &    43.225  &    39.052  &     0.640  \\
SDSS J160913.19+535429.5 &     0.992  &     9.143  &    45.085  &    42.348  &     2.186  \\
SDSS J161156.31+521116.8 &     0.041  &     7.575  &    42.355  &    38.296  &     0.762  \\
SDSS J163709.32+414030.8 &     0.760  &     9.329  &    44.918  &    41.300  &     1.040  \\
SDSS J163856.53+433512.5 &     0.339  &     9.233  &    44.513  &    41.345  &     2.290  \\
SDSS J164108.72+433612.3 &     1.068  &     9.487  &    45.192  &    41.801  &     1.715  \\
SDSS J164224.30+444509.8 &     0.368  &     7.878  &    45.100  &    40.138  &     1.045  \\
SDSS J164258.81+394836.9 &     0.593  &     9.060  &    45.730  &    44.026  &     3.334  \\
SDSS J164829.25+410405.5 &     0.852  &     8.436  &    44.720  &    42.872  &     3.118  \\
SDSS J165005.47+414032.4 &     0.585  &     8.906  &    45.174  &    42.461  &     2.929  \\
SDSS J165641.51+372639.7 &     0.484  &     8.211  &    44.326  &    39.879  &     0.896  \\
SDSS J165819.54+623823.1 &     0.703  &     8.603  &    44.910  &    40.667  &     1.069  \\
SDSS J170306.09+615244.3 &     1.919  &     9.484  &    45.710  &    43.280  &     2.378  \\
SDSS J171300.69+572530.2 &     0.360  &     8.250  &    44.397  &    39.821  &     0.698  \\
SDSS J171936.70+604748.1 &     1.076  &     8.059  &    45.290  &    41.777  &     1.849  \\
SDSS J172051.15+620944.3 &     1.010  &     9.065  &    45.473  &    41.831  &     1.919  \\
SDSS J172206.03+565451.6 &     0.425  &     7.473  &    45.096  &    41.484  &     2.323  \\
SDSS J172659.45+594017.8 &     0.725  &     9.085  &    44.742  &    41.834  &     2.626  \\
SDSS J172750.70+575112.8 &     0.592  &     8.841  &    44.856  &    40.472  &     1.809  \\
SDSS J220908.24-005558.9 &     0.530  &     8.748  &    45.261  &    41.423  &     2.269  \\
SDSS J221542.29-003609.7 &     0.099  &     7.578  &    43.859  &    38.762  &     0.564  \\
SDSS J231845.81-000754.8 &     0.866  &     8.545  &    45.461  &    41.285  &     1.885  \\
SDSS J233624.04+000246.0 &     1.095  &     9.103  &    45.591  &    42.034  &     1.751  \\
SDSS J235156.12-010913.3 &     0.174  &     9.093  &    45.079  &    41.686  &     2.404  \\
\enddata
\end{deluxetable}


\begin{deluxetable}{lccccccr}
\tabletypesize{\scriptsize} \tablewidth{0pt} \tablecaption{The
partial correlation test for the fundamental plane
correlation\label{tbl-2}}

\tablehead{ \colhead{X}           & \colhead{Y} & \colhead{Z} &
\colhead{Type} & \colhead{Number}  & \colhead{$\tau$} &
\colhead{$\sigma$} & \colhead{P$_{null}$}} \startdata
logLx&logLr&logD&radio loud&76&    0.305  &    0.0738  &  3.591E-05 \\
logLx&logLr&logD&radio quiet&39&    0.367  &    0.1059  &  5.295E-04 \\
logLx&logLr&logD&total&115&    0.279  &    0.0562  &  7.009E-07 \\
logLx&logLr&logM&radio loud&76&    0.502  &    0.0550  &  $<$1.000E-10 \\
logLx&logLr&logM&radio quiet&39&    0.623  &    0.0929  &  $<$1.000E-10 \\
logLx&logLr&logM&total&115&    0.491  &    0.0461  &  $<$1.000E-10 \\
logM&logLr&logLx&radio loud&76&    0.131  &    0.0653  &  4.475E-02 \\
logM&logLr&logLx&radio quiet&39&    0.179  &    0.0690  &  9.270E-03 \\
logM&logLr&logLx&total&115&    0.182  &    0.0492  &  2.224E-04 \\
logM&logLx&logLr&radio loud&76&   0.202  &    0.0691  &  3.473E-03 \\
logM&logLx&logLr&radio quiet&39&    0.297  &    0.0960  &  2.013E-03 \\
logM&logLx&logLr&total&115&    0.290  &    0.0531  &  4.608E-08 \\
log(Lx$/L_{Edd}$)&log(Lr$/L_{Edd}$)&logD&radio loud&76&    0.386  &    0.0633  &  1.099E-09 \\
log(Lx$/L_{Edd}$)&log(Lr$/L_{Edd}$)&logD&radio quiet&39&    0.526  &    0.1085  &  1.251E-06 \\
log(Lx$/L_{Edd}$)&log(Lr$/L_{Edd}$)&logD&total&115&    0.307  &    0.0482  &  2.028E-10 \\
log(Lx$/L_{Edd}$)&log(Lr$/L_{Edd}$)&logM&radio loud&76&    0.497  &    0.0515  &  $<$1.000E-10 \\
log(Lx$/L_{Edd}$)&log(Lr$/L_{Edd}$)&logM&radio quiet&39&    0.587  &    0.0777  &  $<$1.000E-10 \\
log(Lx$/L_{Edd}$)&log(Lr$/L_{Edd}$)&logM&total&115&    0.444  &    0.0434  &  $<$1.000E-10 \\
logM&log(Lr$/L_{Edd}$)&log(Lx$/L_{Edd}$)&radio loud&76&    0.143  &    0.0646  &  2.695E-02 \\
logM&log(Lr$/L_{Edd}$)&log(Lx$/L_{Edd}$)&radio quiet&39&   -0.038  &    0.0962  &  6.925E-01 \\
logM&log(Lr$/L_{Edd}$)&log(Lx$/L_{Edd}$)&total&115&    0.211  &    0.0518  &  4.554E-05 \\
logM&log(Lx$/L_{Edd}$)&log(Lr$/L_{Edd}$)&radio loud&76&   -0.269  &    0.0672  &  6.421E-05 \\
logM&log(Lx$/L_{Edd}$)&log(Lr$/L_{Edd}$)&radio quiet&39&   -0.049  &    0.0992  &  6.237E-01 \\
logM&log(Lx$/L_{Edd}$)&log(Lr$/L_{Edd}$)&total&115&   -0.217  &    0.0567  &  1.295E-04 \\
\enddata
\end{deluxetable}

\begin{deluxetable}{lccccr}
\tabletypesize{\scriptsize} \tablewidth{0pt} \tablecaption{The
derived fundamental plane relation \label{tbl-3}}

\tablehead{ \colhead{Subsample}           & \colhead{Number} &
\colhead{$\xi_{RX}$} & \colhead{$\xi_{RM}$} & \colhead{Const.}  &
\colhead{$\sigma_r$}} \startdata
Total&115&$1.33\pm 0.15$&$0.30\pm 0.18$&$-0.40\pm 0.14$&0.89 \\
Radio loud&76&$1.39\pm 0.17$&$0.17\pm 0.21$&$-0.17\pm 0.21$&0.77 \\
Radio quiet&39&$0.85\pm 0.10$&$0.12\pm 0.13$&$-0.77\pm 0.07$&0.38 \\
Merloni et al. (2003)&-&$0.60\pm 0.11$&$0.78^{+0.11}_{-0.09}$&$7.33^{+4.05}_{-4.07}$&0.88 \\
\enddata
\end{deluxetable}


\begin{deluxetable}{lccccccr}
\tabletypesize{\scriptsize} \tablewidth{0pt} \tablecaption{The
partial correlation test for the beaming effect\label{table4}}

\tablehead{ \colhead{X}           & \colhead{Y} & \colhead{Z} &
\colhead{Type} & \colhead{Number}  & \colhead{$\tau$} &
\colhead{$\sigma$} & \colhead{P$_{null}$}} \startdata
$\delta logL_r$&log r&log$L_r$&radio loud&76&    0.514  &    0.0794  &  $<$1.000E-10 \\
$\delta logL_r$&log r&log$L_r$&radio quiet&39&    0.437  &    0.0778  &  1.924E-08 \\
$\delta logL_r$&log r&log$L_r$&total&115&    0.590  &    0.0589  &  $<$1.000E-10 \\
$\delta log\frac{L_r}{L_{Edd}}$&log r&log$L_r$&radio loud&76&    0.504  &    0.0938  &  7.676E-08 \\
$\delta log\frac{L_r}{L_{Edd}}$&log r&log$L_r$&radio quiet&39&    0.369  &    0.0792  &  3.136E-06 \\
$\delta log\frac{L_r}{L_{Edd}}$&log r&log$L_r$&total&115&    0.576  &    0.0616  &  $<$1.000E-10 \\
\enddata

\end{deluxetable}

\clearpage



\begin{thebibliography}{}
\bibitem[Akritas \& Siebert (1996)]{as96} Akritas, M. G., \& Siebert, J. 1996 \mnras, 278, 919
\bibitem[Anderson et al.(2003)]{a03} Anderson, S. F., Voges, W., Margon, B. et al. 2003, \aj,
    126, 2209
\bibitem[Becker et al. (2003)]{becker03} Becker, R. H., Helfand, D. J., White, R.
L. et al. 2003, yCat., 8071, 0
\bibitem[Boroson \& Green (1992)]{bg92} Boroson, T. A., \& Green, R. F. 1992, \apjs, 80, 109
\bibitem[Bregman (2005)]{b05} Bregman, J. N. 2005, submitted to \apjl (astro-ph/0511368)
\bibitem[Brunner et al. (1997)]{bs97} Brunner, H. et al. 1997, \aap, 326, 885
\bibitem[Falcke et al. (2004)]{fkm04} Falcke, H., K$\ddot{o}$rding,
E., \& Markoff, S. 2004, A\&A, 414, 895
\bibitem[Fender et al. (2003)]{fgj03} Fender, R. P., Gallo, E., \& Jonker, P. G. 2003,
MNRAS, 343, L99
\bibitem[Gallo et al. (2003)]{gfp03} Gallo, E., Fender, R. P., \& Pooley, G. G. 2003,
MNRAS, 344,60
\bibitem[Heinz \& Sunyaev (2003)]{hs03} Heinz, S., \& Sunyaev, R.
A. 2003, MNRAS, 343, L59
\bibitem[Heinz (2004)]{h04} Heinz, S., 2004, MNRAS, 355, 835
\bibitem[Isobe et al. (1990)]{i90} Isobe, T., Feigelson, E. D.,
Akritas, M. G., \& Babu, G. J. 1990, ApJ, 364, 104
\bibitem[Kaspi et al. (2000)]{k00} Kaspi, S., Smith, P. S.,
Netzer, H. et al. 2000, ApJ, 533, 631
\bibitem[Kaspi et al. (2005)]{k05} Kaspi, S., Maoz, D.,
Netzer, H. et al. 2005, ApJ, 629, 61
\bibitem[Kellermann et al. (1989)]{k89} Kellermann, K. I., Sramek, R., Schmidt, M. et al.
1989, \aj, 98, 1195
\bibitem[Krolik (1999)]{k99} Krolik, J. H. 1999, Active galactic nuclei: from the
central black hole to the galactic environment (Princeton:
Princeton University press)
\bibitem[McLure \& Jarvis (2002)]{mj02} McLure, R. J., \& Jarvis,
M. J., 2002, \mnras, 337, 109
\bibitem[Merloni \& Fabian (2002)]{mf02} Merloni, A., \& Fabian, A. C. 2002, \mnras, 332, 165
\bibitem[Merloni et al. (2003)]{mhd03} Merloni, A., Heinz, S., \& Di Matteo, T. 2003, MNRAS,
345, 1057
\bibitem[Narayan, Mahadevan \& Quataert (1998)]{n98}Narayan, R., Mahadevan, R. Quataert, E.,  1998, In Theory
of Black Hole Accretion Disks, eds Abramowicz M., Bjornsson G., \& Pringle J.,
Cambridge University Press
\bibitem[Narayan \& Yi (1994)]{ny94} Narayan, R., \& Yi, I. 1994,
\apj, 428, L13
\bibitem[Shakura \& Sunyaev (1973)]{ss73} Shakura, N. I., \&
Sunyaev, R. A. 1973, \aap, 24, 337
\bibitem[White et al. (1997)]{white97} White, R. L., Becker, R. H.,
Helfand, D. J., Gregg, M. D. 1997, \apj, 475, 479
\bibitem[Vestergaard \& Wikes (2001)]{vw01} Vestergaard, M., \&
Wilkes, B. J. 2001, \apjs, 134, 1
\bibitem[Wu et al. (2004)]{w04} Wu, X-B., Wang, R., Kong, M. Z.
et al. 2004, \aap, 424, 793
\bibitem[Yuan et al. (2005)]{y05} Yuan, F., Cui, W., \& Narayan, R. 2005, \apj, 620, 905
\bibitem[Yuan \& Cui (2005)]{yc05} Yuan, F., \& Cui, W. 2005, ApJ, 629, 408
\bibitem[Yuan et al. (1998)]{y98}
Yuan, W., Brinkmann, W., Siebert, J., Voges, W. 1998, \aap, 330, 108


\end{thebibliography}
\end{document}